\documentstyle[12pt,axodraw,epsfig,prc,aps]{revtex}
\newcommand{\fslash}[1]{\mbox{$\!\not\!#1$}}
\begin{document} 
\draft
\title{S-wave $\pi - \pi$ Scattering Lengths in the SU(2) NJL Model \\
Beyond Mean-field Approximation }

\author{Mei Huang$^{1}$, Pengfei Zhuang$^{2,3}$, Weiqin Chao$^{2,1}$ \\ 
{ \small $^1$ Institute of High Energy Physics, Chinese Sciences Academy, Beijing 100039, China \\
  $^2$ China Center of Advanced Science and Technology, World Laboratory,Beijing 100080, China \\
  $^3$ Physics Department, Tsinghua University, Beijing 100084, China\\}}
\date{\today}

\maketitle

\vskip 1cm

\begin{abstract}
The S-wave $\pi - \pi$ scattering lengths $a_0$ and $a_2$
are calculated to the lowest
order of $1/N_c$ expansion in the general 
framework of SU(2) NJL model beyond mean-field approximation. 
It is shown that using the universal curve of  
$a_0$ and $a_2$ 
the four NJL parameters, i.e.,
the current quark mass $m_0$, the four fermion coupling constant $G$,
the quark momentum cut-off $\Lambda_f$ and the meson momentum cut-off
$\Lambda_b$, also the S-wave $\pi - \pi$ scattering lengths
in NJL model can be uniquely determined.

\end{abstract}

\newpage
\section{Introduction}

Pion-pion scattering at threshold became more interesting recently,
because it could provide a test to the mechanism of chiral dynamics.

The Nambu-Jona-Lasinio (NJL) model \cite{NJL} has been regarded 
as a cornerstone
to understand the chiral dynamics by assuming that the spontaneous
chiral symmetry breaking be triggered by the large quark 
condensate in the vacuum. 
And it is described through two Schwinger-Dyson (SD) equations  
for quark propagator and meson propagator respectively \cite{NJL1} - \cite{NJL3}. 
The S-wave $\pi - \pi$ scattering lengths
$a_0$ and $a_2$
have been investigated to the leading order of $1/N_c$ expansion 
(where $N_c$ is the number
of color degrees of freedom) in the mean-field approximation, 
i.e.,Hartree plus RPA,
of SU(2) NJL model \cite{mean1} - \cite{zhuang}.  We know that at the mean-field
approximation level, the two SD equations are not fully coupled,
i.e., the solution of the meson SD equation has no feedback to the quark
propagator, which induces the loss of the information of meson modes
\cite{Ann} \cite{huang}.
Therefore, it is necessary to go beyond mean-field approximation.

In this paper, we calculate the S-wave pion-pion
scattering lengths $a_0$ and $a_2$ in the general 
framework of SU(2) NJL model including current quark 
mass explicitly \cite{huang}, which is 
based on a chirally symmetric self-consistent scheme
considering meson cloud contributions \cite{Ann}.
There are four parameters to be fixed, 
i.e., the current quark mass $m_0$, 
the four fermion coupling constant $G$,
the quark momentum cut-off $\Lambda_f$ and the meson momentum cut-off
$\Lambda_b$. 
 
It is very difficult to extract precise values of the two scattering amplitudes 
in the experimental analysis \cite{exp1}.
However, from either forward dispersion relations or from the Roy equations, 
one can find that $a_0$ and $a_2$ are constrained 
to lie on a universal curve \cite{shaw}.
We try to find an appropriate series of NJL parameters 
by using the new universal scattering length 
relation newly evaluated by M. G. Olsson in \cite{olsson}
together with the two observables $m_{\pi}=139~{\rm MeV}$ and 
$f_{\pi}=92.4 ~{\rm MeV}$.

\section{SU(2) NJL Model Beyond Mean-field Approximation}
~~The two-flavor NJL model is defined through the Lagrangian density,
\begin{eqnarray}
\label{lagr}
{\cal L} = \bar{\psi}(i\gamma^{\mu}\partial_{\mu}-m_0)\psi + 
  G[(\bar{\psi}\psi)^2 + (\bar{\psi}i\gamma_5{\bf {\vec \tau}}\psi)^2 ],
\end{eqnarray}
where $G$ is the effective coupling constant with dimension ${\rm GeV}^{-2}$, 
$m_0$ the current quark mass,
and $\psi, \bar{\psi}$ quark fields with flavor, colour and spinor indices
suppressed, assuming isospin degeneracy of the 
$u$ and $d$ quarks.

We first briefly review the general scheme of the NJL model beyond mean-field 
approximation \cite{huang}.
Including current quark mass explicitly, 
the quark self-energy $m$ expanded to $O(1/N_c)$ order can be expressed as
\begin{eqnarray}
\label{gap}
m = m_0 + m_H +\delta m, 
\end{eqnarray}
where $m_H$ and $\delta m$ are the leading $O(1)$ and subleading  
$O(1/N_c)$ contributions shown in Fig. 1. The solid 
lines in Fig. 1
indicate quark propagator $S(p)=1/({\fslash p}-m)$ with full $m$.
The quark condensate $<{\bar q}q>$ is a one-loop quark integral 
\begin{eqnarray}
\label{cond}
<{\bar q}q>=\frac{1}{N_f}<{\bar \psi} \psi>
           =-4iN_c m \int\frac{d^4p}{(2\pi)^4}\frac{1}{p^2-m^2}.
\end{eqnarray}
The corresponding meson propagator $D_{M}(k)$ (M means 
$\pi$ or $\sigma$) has the form
\begin{eqnarray}
\label{prog}
-{\rm i} D_{M}(k) & = & \frac{2iG}{1-2G \Pi_{M}(k)}, \nonumber \\
\Pi_{M}(k) & = & \Pi_{M}^{(RPA)}(k)+\delta\Pi_{M}^{(b)}(k)+
             \delta\Pi_{M}^{(c)}(k)+\delta\Pi_{M}^{(d)}(k),
\end{eqnarray}
where $\Pi_{M}$ is the meson polarization function, which includes
the leading
order $\Pi_{M}^{(RPA)}$ and subleading order $\delta\Pi_{M}^{(b,c,d)}$, 
shown in Fig. 1. The above constituent quark mass and the meson propagator,
namely the Eqs. ~(\ref{gap}) and (\ref{prog}), or the Feynman diagrams
in Fig. 1,~ form a self-consistent description of the SU(2) NJL model to the
subleading order of $1/N_c$ expansion, which is different from the earlier
calculations \cite{NJL1} - \cite{schu} where only the mean-field quark
mass $m_H$ and the Random-Phase-Approximation (RPA) meson polarization function 
$\Pi_{M}^{(RPA)}$ are considered. 

The meson mass $m_{M}$ satisfies the total meson propagator's pole condition
\begin{eqnarray}
\label{pole}
1-2G\Pi_{M}(k^2=m_{M}^2)=0,
\end{eqnarray}
and the meson-quark coupling constant $g_{M qq}$ is determined by the residue at the pole
\begin{eqnarray}
\label{couple}
g_{M qq}^{-2} = (\partial \Pi_{M}(k)/\partial k^2)^{-1}|_{k^2=m_{M}^2}.
\end{eqnarray}
Another important quantity in the meson sector is the pion decay constant $f_{\pi}$ 
which 
generally satisfies 
\begin{eqnarray}
\label{onshell}
\frac{m_{\pi}^2f_{\pi}}{g_{\pi qq}} = \frac{m_0}{2G}.
\end{eqnarray}
In the chiral limit, $f_{\pi}$ satisfies the Goldberger-Treiman relation 
$f_{\pi}(k)g_{\pi qq}(k)=m$.

\section{$\pi - \pi$ scattering at threshold and the universal curve}

Now we turn to the calculation of the
S-wave $\pi - \pi$ scattering length $a_0$ and $a_2$. 
The invariant scattering amplitude can be generally written as \cite{shaw}
\begin{eqnarray}
T_{ab,cd}=A(s,t,u)\delta_{ab} \delta_{cd}
          + B(s,t,u)\delta_{ac} \delta_{bd} + C(s,t,u) \delta_{ad} \delta_{bc},
\end{eqnarray}
where $a, b$ and $c, d$ are the isospin labels of the initial and final states 
respectively, and $s, t$ and $u$ are the Mandelstam variables, $s=(p_a+p_b)^2$,
$t=(p_a-p_c)^2$, $u=(p_a-p_d)^2$, 
where $p_a \sim p_d$ are initial and final isospin momentum.
Using perfect crossing symmetry, one can 
project out isospin amplitudes
\begin{eqnarray}
A_0=3A+B+C, \ \ \ A_1=B-C, \ \ \ A_2=B+C.
\end{eqnarray}
In the limit of scattering at threshold, ${\sqrt s}=2m_{\pi}$, t=u=0, the S-wave 
scattering lengths $a_{I}$ (given in units of $m_{\pi}^{-1})$ are:
\begin{eqnarray}
a_{I}=\frac{1}{32 \pi}A_I(s=4m_{\pi}^2, t=0, u=0), \ \ I=0,2.
\end{eqnarray}

It is known that the two scattering lengths $a_0$ and $a_2$
at threshold are not independent, 
they are constrained to lie on a universal curve \cite{shaw} \cite{olsson}:
\begin{equation}
2 a_0 -5 a_2=\frac{12}{\pi}\int_0^{\infty}\frac{d q}{q (1+q^2)} {\rm Im} A_{I_t=1}(q),
\end{equation}
where $q$ is the c.m. momentum in units of $m_{\pi}$.
And in \cite{olsson}, M. G. Olsson evaluated the new universal relation as:
\begin{eqnarray}
\label{curve}
2 (a_0 - \frac{4}{3 \pi} a_0^2) -5 (a_2- \frac{4}{3 \pi} a_2^2)
               = L_0,
\end{eqnarray}
\begin{eqnarray}
\label{l0}
L_0=\frac{12}{\pi}\int_0^{\infty}\frac{d q}{q (1+q^2)} {\rm Im} A^{(0)}_{I_t=1}(q),
\end{eqnarray}
The integrand of $L_0=0.58 \pm 0.015~ m_{\pi}^{-1}$ in \cite{olsson} 
includes the dominant contribution of $\rho(770)$ and 
other small contributions of higher resonances.   
    
We will calculate the scattering lengths to the 
lowest order in $1/N_c$ expansion of the process $\pi \pi \rightarrow \pi \pi$
in the general framework of SU(2) NJL model 
beyond mean-field approximation.
The Feynman diagrams 
include box and $\sigma-$
exchange diagrams, which are the same as those  
in \cite{schu}. The only difference is that in \cite{schu}
quark mass $m_H$ is in the mean-field approximation and 
meson polarization function $\Pi_M^{(RPA)}$ in RPA, while in our calculations,
quark mass $m$ and meson polarization function $\Pi_M$ 
include the subleading order contributions. 
 
To keep the lowest order of the $\sigma$ exchange diagrams, 
the internal $\sigma$ propagator  
should be in leading $O(1/N_c)$ order,
which has the same form as that in RPA $\Pi_M^{(RPA)}$ \cite{huang}.
So what we need to do is using the expression formulae given in \cite{schu} 
directly,
and replacing the quantities in the mean-field approximation with those beyond
mean-field approximation, i.e.,
the external pion propagator becomes the total propagator Eq. (\ref{prog}),
the pole for quark propagator becomes the total quark self-energy $m$ defined in 
Eq. (\ref{gap}), and the coupling constant 
$g_{\pi qq}$ now is the one expressed in Eq. (\ref{couple}). 

\section{Numerical Results}
For the numerical calculations, we adopt the external
momentum expansion method as discussed in detail in \cite{huang}.
We introduce a quark momentum cut-off $\Lambda_f$ in 
Pauli-Villars regularization 
and a meson momentum cut-off $\Lambda_b$ in covariant regularization 
for the divergent momentum integrals. Using only the two experimental 
observables 
$m_{\pi}=139 ~{\rm MeV}$ and
$f_{\pi}=92.4 ~{\rm MeV}$,
one can not give fixed values of the four parameters in the model, namely the 
current quark mass $m_0$,
coupling constant $G$, and the two momentum cuts $\Lambda_f$ and $\Lambda_b$. 
We introduce
one more free parameter $z=\Lambda_b / \Lambda_f$, 
which characterizes the meson cloud 
contributions. Especially,
in the limit of $z=0$, the model goes back to the
mean-field approximation automatically.

For each $z$, with quark mass $m$ changing from $200 ~{\rm MeV}$ to 
$1200 ~{\rm MeV}$, 
we solve the three
equations (\ref{gap}), (\ref{pole}) and (\ref{onshell}) to get 
a series of $\Lambda_f$,
$m_0$ and $G$, and then calculate other quantities like 
$<{\bar q}q >$ and $g_{\pi qq}$.   
Finally, we can calculate the 
S-wave $\pi - \pi $ scattering lengths $a_0$ and $a_2$ as functions of $m$. 

We show in Fig. 2 $a_0$ as a function of $m$ for different $z$.
It is seen that for each $z$, there is a minimum value of $a_0$,  
around which there is a plateau that $a_0$ changes slowly with $m$.
We have pointed out in \cite{huang}, that for each $z$, there 
is also a plateau around the minimum 
of quark condensate varying slowly with $m$.
The two $m$ regions corresponding to 
the $a_0$ plateau and quark condensate plateau almost coincide.

And in Fig. 3, $a_2$ as a function of $a_0$ for different $z$ is shown 
in the $(a_0,a_2)~ plane$.
It can be seen that in the $(a_0,a_2)~ plane$, 
there is a turning-point of the NJL $a_2(a_0)$ curve for each $z$, 
which corresponds to the minimum 
value of $a_0$ for each $z$. 
It is noticeable that all the turning-points for different $z$ 
in the $(a_0,a_2)~plane$ are on a line!
Remembering that $a_0$ and $a_2$ are constrained 
to lie on a universal curve Eqs. (\ref{curve}) 
with $L_0=0.58 \pm 0.015~ m_{\pi}^{-1}$, 
we show this curve as double solid lines in the $(a_0,a_2)~plane$.
It can be seen that the turning-point
of the NJL $a_2(a_0)$ curve for $z=0.7$ falls on the universal curve.  
This turning-point 
$(a_0=0.1796~ m_{\pi}^{-1}$, $a_2=-0.0489~ m_{\pi}^{-1})$ is shown 
as a solid circle
in the $(a_0,a_2)~plane$.

To compare our results with other theoretical predictions, we also plot 
all the predictions (in units of $m_{\pi}^{-1}$) in Fig. 3:
1), The Weinberg values 
$(a_0=0.158, a_2=-0.045)$ \cite{weinberg}; 2), The chiral 
perturbation theory at one-loop (ChPT 1loop) \cite{gasser} predicted 
$(a_0=0.200, a_2=-0.043)$; 3), The chiral 
perturbation theory at two-loop (ChPT 2loop) \cite{bijnens} predicted 
$(a_0=0.217, a_2=-0.0413)$; 4), Olsson's dispersion-relation
constraint of the heavy baryon  chiral 
perturbation theory \cite{bernard} results (HBChPT Olsson) \cite{olsson}
$(a_0=0.235, a_2=-0.031)$. 
It can be seen that the turning-point
of the NJL $a_2(a_0)$ curve for
$z=0$, i.e., $(a_0=0.1576, a_2= -0.0446)$ shown as a $\times$
coincides with the Weinberg values shown as a star. 

\section{Conclusions and Discussions}
Our results show that, in the framework of SU(2) NJL model, 
there is a turning-point of the $a_2(a_0)$ curve for each $z$ 
in the $(a_0, a_2)~plane$, 
and the turning-point 
corresponds to the region of the quark-condensate plateau;
And the turning-point for $z=0$ is almost the same as the 
Weinberg values;
Moreover, we find that all the turning-points in the $(a_0, a_2)~plane$
are on a line, which has a intersecting point
with the universal curve at the turning-point of $z=0.7$.

We then can read the four parameters corresponding to this point:
$m_0=8.22 ~{\rm MeV}$, $\Lambda_f=667.5 ~{\rm MeV}$, 
$\Lambda_b= \Lambda_f \times 0.7 ~{\rm MeV}$, and 
$G \Lambda_f^2= 4.36$. 
And also we can calculate the other corresponding quantities 
at this intersecting point: the constituent 
quark mass $m=400 ~{\rm MeV}$ and the quark condensate
$<{\bar q} q>^{1/3}=223.4 ~{\rm MeV}$.

Also we find that our results of pion-pion scattering lengths
determined by the universal curve, i.e., the turning-point
of $z=0.7$, are smaller than other predictions.
The reasons
maybe lie in that:
a), From \cite{olsson}
we know that the $L_0$ integrand of the universal curve include dominantly
the $\rho$'s contribution and other higher resonances.
Here, we only consider the sigma exchange contributions. If one wants to
investigate $\rho$'s contribution, the
extended NJL model should be used like in Ref. \cite{heyb}; 
b), We only calculate to the lowest order in $1/N_c$ expansion of pion-pion
scattering Feynman diagrams.

\section*{Acknowledgement}
~~The authors would like to thank Dr. Bing-Song Zou
for his help during the work, we also thank S. P. Klevansky
for helpful discussions. This work is supported by NNSF of China
(Nos. 19677102 and 19845001).

\begin{center}
\begin{picture}(500,200)(0,0)
\SetWidth{2}
\CArc(120,50)(15,0,360)
\Vertex(120,35){2.5}
\Text(120,19)[]{$m_H$}
\CArc(200,50)(15,0,360)
\DashCArc(200,70)(15,320,220){4}
\Vertex(200,35){2.5}
\Vertex(189,61){2.5}
\Vertex(210,61){2.5}
\Text(200,19)[]{$\delta m$}
\Text(200,75)[]{$\pi$,$\sigma$}
\end{picture}
\begin{picture}(500,85)(0,0)
\SetWidth{2}
\CArc(105,50)(15,0,360)
\Vertex(90,50){2.5}
\Vertex(120,50){2.5}
\Text(110,15)[]{$\Pi^{(RPA)}_{M}(k)$}
\CArc(180,50)(15,0,360)
\DashLine(180,65)(180,35){4}
\Vertex(165,50){2.5}
\Vertex(180,65){2.5}
\Vertex(180,35){2.5}
\Vertex(195,50){2.5}
\Text(180,15)[]{$\delta \Pi^{(b)}_{M}(k)$}
\Text(180,50)[]{$\pi$,$\sigma$}
\CArc(260,50)(15,0,360)
\DashCArc(260,70)(15,320,220){4}
\Vertex(249,60){2.5}
\Vertex(248,41){2.5}
\Vertex(271,60){2.5}
\Vertex(272,41){2.5}
\Text(260,75)[]{$\pi$,$\sigma$}
\Text(260,15)[]{$\delta \Pi^{(c)}_{M}(k)$}
\SetWidth{2}
\CArc(340,50)(10,0,360)
\CArc(400,50)(10,0,360)
\Vertex(330,50){2.5}
\Vertex(340,60){2.5}
\Vertex(340,40){2.5}
\Vertex(410,50){2.5}
\Vertex(400,60){2.5}
\Vertex(400,40){2.5}
\Text(370,65)[]{$\pi$}
\Text(370,35)[]{$\sigma$}
\Text(370,15)[]{$\delta \Pi^{(d)}_{M}(k)$}
\DashCurve{(340,60)(360,70)(380,70)(400,60)}{3.2}
\DashCurve{(340,40)(360,30)(380,30)(400,40)}{3.2}
\end{picture}
\end{center}
\begin{figure}[ht]
\begin{center}
\caption{Feynman digrams for leading and subleading order quark self-energy 
$m_H$ and $\delta m$, and meson polarization functions $\Pi^{(RPA)}_{M}$ and 
$\delta \Pi^{(b,c,d)}_{M}$. The solid and dashed lines indicate the quark 
internal meson propagators respectively.}
\label{kernelb}
\end{center}
\end{figure}
\newpage
\begin{figure}[ht]
\centerline{\epsfxsize=18cm\epsffile{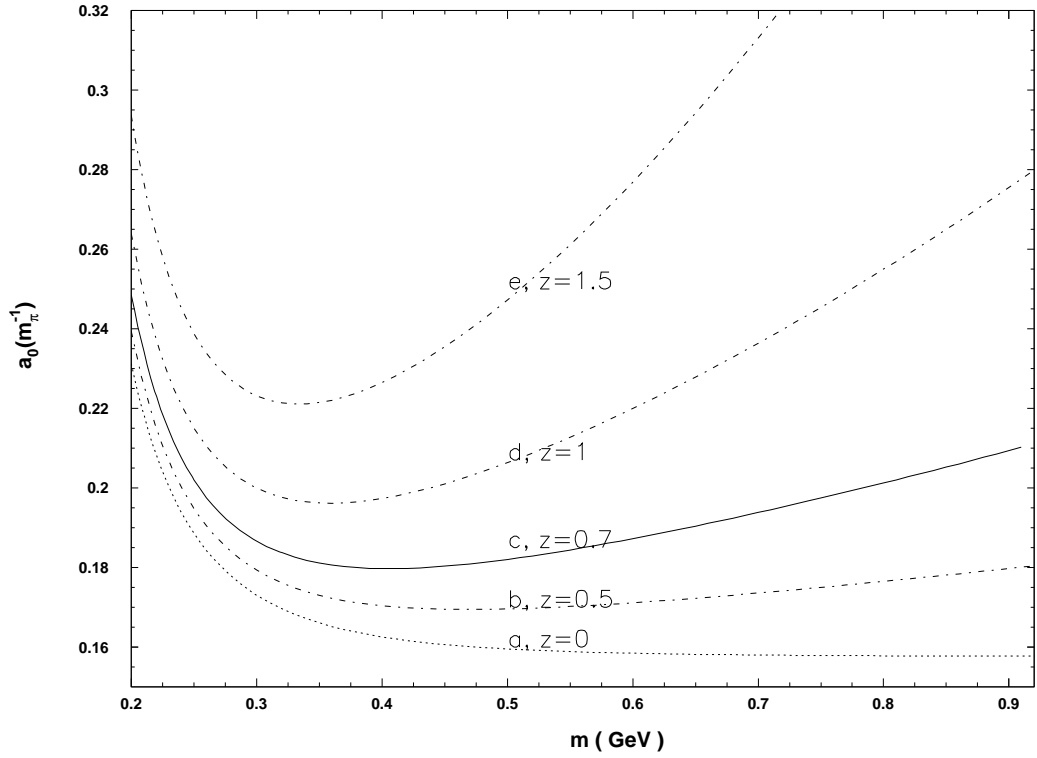}}
\caption
{$a_0$ as a function of $m$ for different $z$.}
\label{a0_fig}
\end{figure}
\newpage
\begin{figure}[ht]
\centerline{\epsfxsize=18cm\epsffile{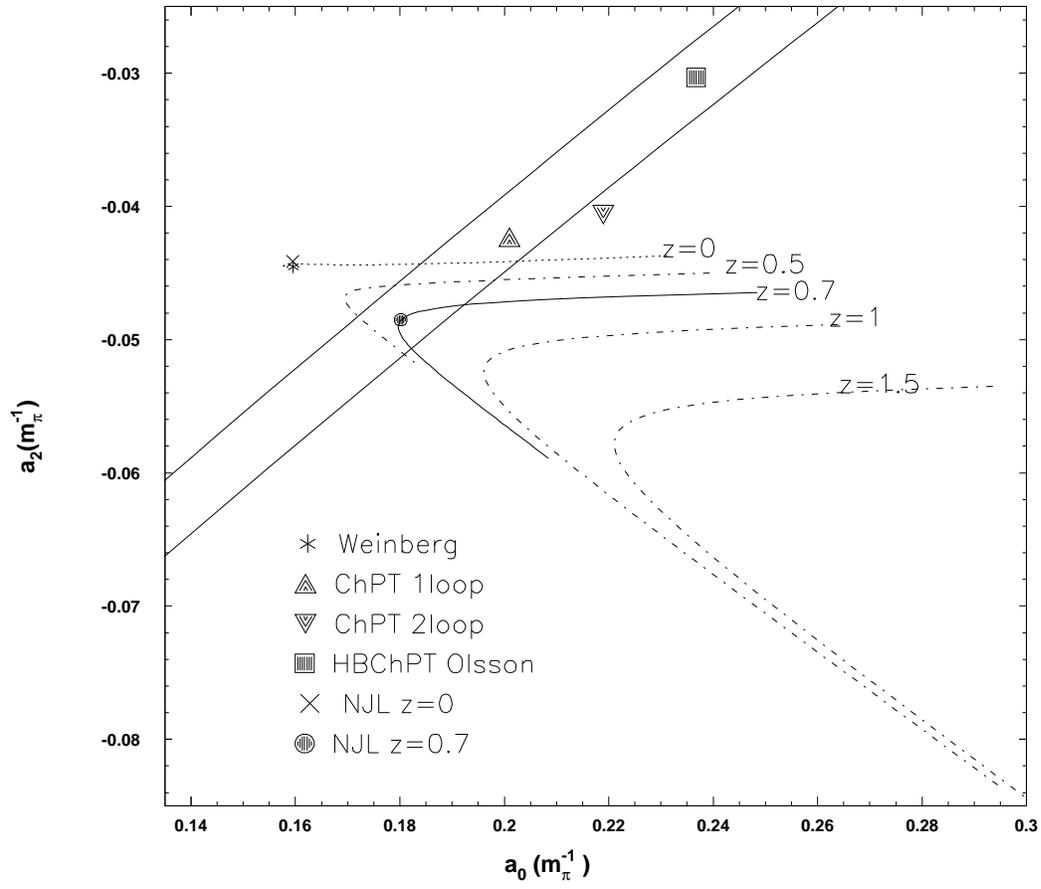}}
\caption
{The $(a_0,a_2)$ plane include 1), $a_2$ as a function of $a_0$ 
for different $z$ i.e. $a_2(a_0)$ curve in NJL model; 
2), the universal curve shown as solid double lines;
3), other theoretical predictions.}
\label{a0a2_fig}
\end{figure}

\end{document}